\documentclass[a4paper]{ESASPStyle}
\newif\ifpdf
\ifx\pdfoutput\undefined
  \pdffalse
  \usepackage{graphicx}
  \DeclareGraphicsExtensions{.eps, .ps, .epsi}
\else
  \pdfoutput=1
  \pdftrue
  \usepackage[pdftex]{graphicx}
  \DeclareGraphicsExtensions{.pdf}
\fi
\newcommand{\figref}[1]{Fig.~\ref{#1}}

\newcommand{\eqref}[1]{Equation~\ref{#1}}

\newcommand{\secref}[1]{Section~\ref{#1}}

\newcommand{\tabref}[1]{Table~\ref{#1}}
\def\note #1]{{\bf #1]}}
\def\newpage{\vfill\eject}

\def\etal{{\it et al.}}
\def\eg{{\it e.g.}}
\def\cf{{\it cf.}}
\def\ie{{\it i.e.}}

\def\Msun{M_{\odot}}

\def\MONS{MONS}

\def\deg{${}^\circ$}

\begin{document}

\title{MONS on the Danish R{\o}mer satellite:
Measuring Oscillations in Nearby Stars}

\author{J{\o}rgen Christensen-Dalsgaard}
  \institute{Teoretisk Astrofysik Center, Danmarks Grundforskningsfond, and \\
  Institut for Fysik og Astronomi, Aarhus Universitet,
  DK 8000 Aarhus C, Denmark}

\maketitle 

\begin{abstract}

MONS (for {\bf M}easuring {\bf O}scillations in {\bf N}earby
{\bf S}tars) is the scientific project on the Danish R{\o}mer
satellite mission,
which is being developed as part of the Danish Small Satellite Programme.
The principal goal is to study solar-like oscillations in
around 20 bright stars, with a precision that in the best cases will
be limited only by the intrinsic stellar `noise'.
The baseline orbit, a so-called Molniya orbit, allows access to essentially
the entire sky during the planned 2-year mission.
The main instrument is a short-focus reflecting telescope with an
aperture of 32~cm, 
making two-colour measurements.
A focused Field Monitor will be used to detect and correct for
possible faint variable stars of substantial amplitude near the main target.
In addition the Field Monitor, and the Star Trackers on the platform,
may be used to observe a broad range of variable phenomena.
The project has concluded the Systems Definition Phase
by a successful review,
and launch is scheduled for the middle of 2005.

\keywords{Stars: structure, oscillations -- asteroseismology --
high-precision photometry}
\end{abstract}

\section{Introduction}

Denmark launched its first national satellite, {\O}rsted, in 1999;
the purpose of this mission is to make very accurate measurements
of the Earth's magnetic field. 
The satellite is still (October 2001) operational and has provided
a number of very important results, including a revised
model of the Earth's magnetic field (Olsen {\etal} 2000).
As a follow-up to the successful development of this mission,
it was decided in 1997 to establish a Danish Small Satellite Programme, 
initially funded for the four-year period 1998 -- 2001.

\begin{figure}[!htb]
\begin{center}
  \resizebox{8.8cm}{!}
  {\includegraphics{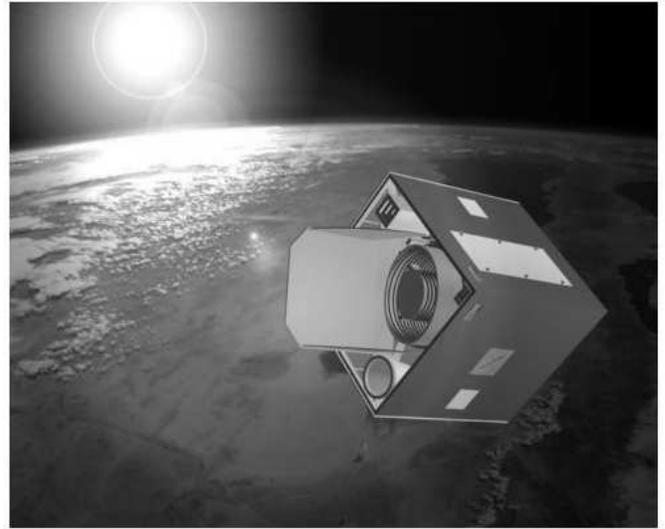}}
\end{center}
\caption[Virgo red and blue spectra]{
\label{fig:inorbit}
{\it Artist's impression of the R{\o}mer satellite in orbit.}
}
\end{figure}

The call for mission ideas within this programme,
and the further more detailed applications
and evaluation, led to the selection of a mission consisting of two
projects:

\begin{itemize}

\item {\bf MONS}, to study solar-like oscillations in other stars

\item {\bf Ballerina}, to detect gamma-ray bursts and observe them
with an X-ray telescope.

\end{itemize}

The proposal, by the committee under the Danish Research Councils 
which is responsible for the programme, to combine these two
missions resulted from a communality of features, including the
orbit, and the expectation that the combined mission would fit
within the constraints of the programme.
It was also stated by the committee that, if the combination turned
our to be infeasible, the MONS project had the highest priority
and should be carried out on its own.
The combined mission was named R{\o}mer, after the Danish
astronomer Ole R{\o}mer%
\footnote{born and initially educated in Aarhus}
(1644 -- 1710); he was the first to measure the finite speed of light
and he made major developments of astronomical instrumentation.

The System Definition Phase of the combined mission was started
early 2000, by a combination of Danish industries and research
institutions. 
An important early goal was to establish whether or not the
combined mission was feasible.
At a mid-term review in October 2000 it was concluded that,
although technically possible, the combined mission was too
complex, and likely too demanding in terms of mass and finance,
to be realistic within the Small Satellite Programme.
Thus it was decided to continue with a mission involving only
the MONS project; however, the name R{\o}mer was kept for this mission.
Figure~\ref{fig:inorbit} illustrates the current configuration of
the satellite.

The goal of the MONS project is to probe the interiors of stars
through observations of oscillations in their intensity and colour.
Thus the mission extends the very successful studies of the
solar interior over the last decade through helioseismology
(for a review see, for example, Christensen-Dalsgaard {\etal} 2000).
The frequencies of oscillation will provide information about
the structure and rotation of the stellar interiors, and hence
allow investigation of effects of stellar evolution and the
physical properties of matter in the stars.
The main targets for the mission are stars that show
oscillations similar to those observed in the Sun.
Since the amplitudes of such oscillations are typically of 
order a few parts per million (ppm) an overriding requirement
on the project is photometric precision;
the design goal is that the instrumental and photon noise
should be substantially below the intrinsic stellar granulation
noise for the brightest objects on the target list.
MONS will observe solar-like oscillations in 15 -- 20 stars.
In addition, high-precision photometry will be carried out on a few
other carefully selected variable stars.
Sufficient frequency resolution requires that each target be
observed almost continuously for about one month.
Also, the primary targets are distributed over the whole sky,
and hence an orbit allowing access to the entire sky is highly desirable.
Given the budget of the programme, launch has to be obtained as a
passenger on a larger mission.
Thus it was decided to keep the satellite within the ASAP-5 restrictions,
which is being established as a {\it de-facto} standard for
such secondary launches.
Thus the dimensions of the satellite are limited to 60 by 60 by 71 cm,
with a total mass below 120 kg.

The primary observations will use the MONS Main Telescope, a strongly
defocussed 32-cm telescope observing in two colours.
However, the satellite will also carry other instruments
permitting a broad range of parallel science.
The MONS Field Monitor has as primary goal to detect and correct
for faint variable stars near the main target;
such stars, within the field of the Main Telescope, might otherwise
corrupt the data.
The Field Monitor has an opening of 5 cm and a field of
$5^\circ \times 5^\circ$, which is observed in focus.
Thus it allows the nearly continuous monitoring of the large number
of other objects in the field, undoubtedly including variable
stars of various types as well as other time-varying astrophysical objects.
Data for an even larger field, although with lower precision, will
be obtained from the Star Trackers, whose main purpose is to
determine the orientation of the satellite.
Two Star Trackers will be included, pointing in the same direction as,
and the opposity direction to, the Main Telescope.
The Star Trackers have a 2.4~cm aperture and cover a field 
with a diameter of $22^\circ$.

The R{\o}mer satellite is predominantly a Danish mission, 
with the scientific and technical responsibility, as well the
overall design and construction of the satellite, being
located in Denmark.
The scientific headquarters of the mission are at University of
Aarhus, the prime industrial contractor is 
the company Terma A/S (see {\tt http://www.terma.com})
and the overall management is carried out at the Danish Space Research
Institute.
However, the project involves very important international contributions.
The design and construction of the Main Telescope will be
carried out by the Australian companies Auspace and Prime Optics,
while the optics and other hardware for the Field Monitor is
provided by Spain and Belgium.
Also, Finland will provide the platform computer,
while contributions to the onboard software development come
from Belgium.
Finally, the scientific preparations for the mission and the utilization
of the data involve the broad MONS Science Consortium, with
around 150 members in 20 countries.

\section{Primary scientific goals}
\label{sec:primgoals}

Observations of stellar oscillations provide information on many aspects of
the stellar interior.  The {\em frequencies\/} of the oscillations depend
on the structure of the star, particularly the distribution of sound speed
and density, and on gas motion and other properties of the stellar
interior.  The {\em amplitudes\/} of the oscillations are determined by the
excitation and damping processes, which may involve turbulence from
convection, opacity variations and magnetic fields.  The best targets are
stars which oscillate in several modes simultaneously.  Each mode has a
slightly different frequency, reflecting spatial variations of the
structure within the star, and the combination places strong constraints on
the internal properties.  (For reviews of such studies see
Brown \& Gilliland 1994; Gautschy \& Saio 1996; Christensen-Dalsgaard 1998;
Gough 1998.)

Stellar oscillations are standing waves which reach deep inside the star,
and they are observed via their effect on the surface.  To a very good
approximation, the surface pattern of a single oscillation mode can be
described by a spherical harmonic, which is characterized by two quantum
numbers: the degree $l$ and the azimuthal order $m$;
$l$~determines the horizontal wavelength,
while $m$ measures the number of nodes around the equator.
In addition, a mode is characterized by the radial order $n$,
which roughly corresponds to the number of nodes in the radial direction.
In practice, for stars apart from the Sun we measure averages over the
stellar surface, thus effectively suppressing modes with comparatively high
degree and hence short wavelength.
In general, stellar observations are dominated by the modes of the lowest
degrees ($l = 0$, 1, 2 and~3).
Fortunately, these are the most useful for probing the
interior because they reach deepest below the stellar surface.

The principal targets for the MONS Main Telescope
are stars with oscillations similar to those observed in the Sun.
These are believed to be excited stochastically by the effects of convection
in the near-surface layers of the star, where the speed of convective 
motion approaches the sound speed, leading to an efficient generation
of sound waves.
These waves couple to the normal modes of the star, leading to 
oscillations which can be observed in, for example, the luminosity
and surface temperature of the star.
This mechanism is expected to cause oscillations in relatively cool
stars with substantial near-surface convection;
these include stars in the phase of central hydrogen fusion with
masses up to around $1.7 \Msun$, $\Msun$ being the mass of the Sun.
Due to the broad-band nature of the mechanism 
most modes in a fairly broad range of frequencies
are expected to be excited, as is indeed observed in the Sun.
This substantially simplifies the identification of the modes
and the analysis of the frequencies.
Unfortunately, the expected amplitudes are extremely small:
a few parts per million (ppm) in relative fluctuations in 
luminosity or surface temperature. 

\begin{figure}[!htb]
\begin{center}
  \resizebox{8.8cm}{!}
  {\includegraphics{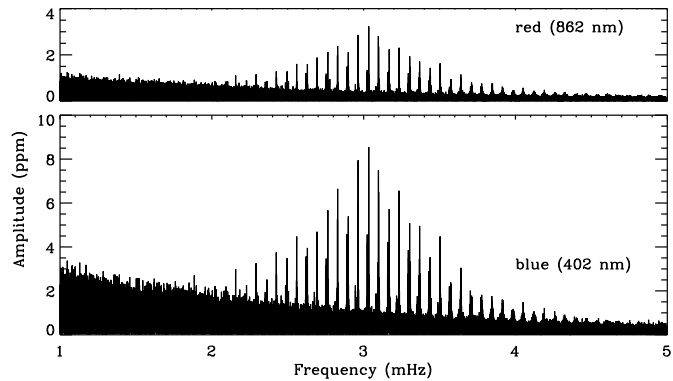}}
\end{center}
\caption[Virgo red and blue spectra]{
\label{fig:VIRGO}
Amplitude spectra of solar oscillations measured in intensity by the VIRGO
instrument on the SOHO spacecraft.  The observations are smoothed and
rescaled here to show the spectrum corresponding to 30 days.  Individual
oscillation modes appear as strong peaks rising above a sloping background,
which arises from random convective motion on the solar surface.  \MONS{}
will produce amplitude spectra of other stars with similar signal-to-noise
levels.  (Adapted from Fr\"ohlich {\etal} 1997.)  }
\end{figure}

\begin{figure}
\begin{center}
  \resizebox{8.8cm}{!}
  {\includegraphics{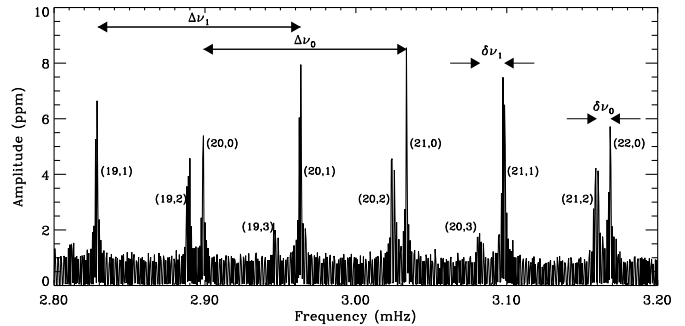}}
\end{center}
\caption[The large and small frequency separations]{\label{fig:schemspec}
Small section of the solar spectrum (lower panel of \figref{fig:VIRGO}),
showing $(n,l)$ values for each mode.  The large and small separations
are indicated.  These measure the average density and core composition,
respectively, and can therefore be used to infer the mass and age of a star.
}
\end{figure}

Figures \ref{fig:VIRGO} and \ref{fig:schemspec} illustrate properties
of the solar oscillation spectrum, observing the Sun in a manner
similar to the MONS observations, in light integrated over the solar disk.
The observed modes are concentrated in a frequency range between 
2 and about 4.5 mHz, corresponding to periods between 8 and 4 minutes.
(Other observations shows that the solar oscillation spectrum extends
to frequencies below 1 mHz.)
These modes are acoustic modes of high radial order which, according
to asymptotic theory, are essentially uniformly spaced in frequency,
as is indeed observed.
It should be noticed that the observed amplitudes are substantially
higher in the blue than in the red wavelength region.
This is the background for using colour as the primary observable
with the MONS Main Telescope: the ratio between blue and red
intensities still shows a substantial oscillation signal,
while certain contributions to the instrumental noise are
strongly reduced.
It should also be noted that the `noise' background in 
Fig.~\ref{fig:VIRGO} is in fact of solar origin:
it arises from the intensity fluctuations caused by granular
motion at the solar surface.

Details of the solar spectrum are illustrated in Fig.~\ref{fig:schemspec}.
It is characterized by the {\it large frequency separation} 
$\Delta \nu_l = \nu_{nl} - \nu_{n-1\,l}$ between modes of adjacent
radial order $n$, corresponding to the almost uniform spacing
shown in Fig.~\ref{fig:VIRGO}, and the 
{\it small frequency separation}
$\delta \nu_l = \nu_{nl} - \nu_{n-1\,l+2}$;
here $\nu_{nl}$ is the cyclic frequency of a mode of radial order $n$
and degree $l$.
The small separation is substantially influenced
by the sound speed in the core of the star and hence is sensitive to the
chemical composition there; since the core composition changes with age, as
hydrogen is converted into helium, $\delta \nu_l$ provides a measure of
the evolutionary state of the star.  On the other hand, $\Delta \nu_l$
depends on the overall properties of the star.

These properties have important consequences for
the diagnostic potentials of the frequencies: by observing
$\Delta\nu_l$ and $\delta\nu_l$ it is in principle possible to
determine both the mass and age of a star.
In practice, the determination is evidently dependent on other 
uncertainties in stellar properties.
These issues are discussed in more detail by Gough (these proceedings)
and Monteiro, Christensen-Dalsgaard \& Thompson (these proceedings).

Detailed observations of accurate frequencies offer far more information
about the stellar interior, allowing inferences beyond the basic parameters
of the star such as mass and age.  For example, the location of the base of
the convective envelope introduces an oscillatory signal in the oscillation
frequencies which may be used to constrain the depth
of the convection zone and the properties of the region below it
(Monteiro, Christensen-Dalsgaard \& Thompson 2000;
Monteiro {\etal}, these proceedings).
Also, P\'erez Hern\'andez \& Christensen-Dalsgaard (1998) found that
the ionization of helium affects the sound speed and hence the oscillation
frequencies of stellar models in a manner that is in principle observable;
this may be used to determine stellar abundances of helium, a quantity that
is otherwise only poorly known.
With sufficiently good data it may be possible to carry out inversions
for stellar structure, at least in parts of the star or under
additional assumptions
(see Basu, Christensen-Dalsgaard \& Thompson, these proceedings;
Roxburgh, these proceedings;
Thompson \& Christensen-Dalsgaard, these proceedings).

Estimates of the amplitudes expected for solar-like oscillations
have been made, assuming stochastic excitation by convection
and calibrating against the solar amplitudes
({\eg} Christensen-Dalsgaard \& Frandsen 1983;
Houdek {\etal} 1999).
Kjeldsen \& Bedding (1995) pointed out that the results roughly
indicated that the amplitudes were proportional to the ratio
$L/M$ between the stellar surface luminosity and mass.
Recent observations of solar-like oscillations
(Kjeldsen {\etal} 1995; Marti\'c {\etal} 1999;
Bedding {\etal} 2001; Bouchy {\etal} 2001)
indicate that the amplitudes increase less rapidly with increasing 
effective temperature, but show the predicted increase with
luminosity at solar effective temperature.
Nevertheless, these observations provide confidence that MONS will
indeed be able to make detailed observations of oscillations in
a substantial range of stars.

Although the main emphasis will be on solar-like oscillations,
the MONS Main Telescope will also be used to observe
a few stars known from ground-based observations to pulsate.
Theory suggests that in these many more modes may be excited
than currently observed, likely at amplitudes rendering them 
undetectable from the ground.
Important examples are the $\delta$ Scuti stars and the 
$\beta$ Cephei stars, which together cover an extended
range in mass, and hence interior properties, along the core
hydrogen-burning main sequence.
If such low-amplitude modes can indeed be detected, the MONS observations may 
provide very valuable information about the properties of the stars.
Additional data on such `classical' pulsating stars will be
obtained with the R{\o}mer Star Trackers and the
MONS Field Monitor ({\cf} \secref{sec:parallel}).

In addition to stellar structure, the frequencies of oscillation
depend on the rotation of the star.
This gives rise to a frequency splitting according to $m$ which,
to lowest order,
can be roughly written as
\begin{equation}
\nu_{n l m} 
= \nu_{n l 0} + m { \langle \Omega \rangle_{nl} \over 2 \pi } \; ,
\label{rotsplit}
\end{equation}
where $\langle \Omega \rangle_{nl}$ is an average of the angular
velocity weighted by the structure of the given mode.  When only low-degree
modes are observed, as will be the case for \MONS{}, these averages all
extend over most of the star.  However, even such limited information will
give some indication of the variation of rotation with depth in the star,
particularly if it is combined with measurement of the surface rotation via
intensity variations induced by spots, which \MONS{} will easily detect.
Measurement of the rotational splitting requires observations over a period
that somewhat exceeds the rotation period.  Thus \MONS{} will be able to
measure the splitting for stars rotating at least as fast as the Sun, with
rotation periods below 25 days%
\footnote{It should be noted that for relatively rapid rotation
the effects on the frequencies are substantially more complicated;
see, for example, Soufi {\etal} (1998).}.

\section{Payload}
\label{sec:payload}

The primary requirement on the payload and platform is to allow
measurements of oscillations in intensity or colour at the sub-ppm level
over a one-month observing sequence, in the frequency range relevant
to stellar oscillations.
To achieve a sufficiently low photon noise, and hence sufficiently
large number of photons, with current CCD detectors strong defocussing
is required.
As a result, the Main Telescope cannot distinguish between the 
bright primary target and possible faint near-by stars varying
with substantial amplitudes;
to correct for this, the payload includes a focussed Field Monitor
which can detect such variables and correct for their influence.

\subsection{MONS Main Telescope}
\label{sec:MMT}

The telescope aperture,
together with the optical and detector efficiency,
determines the photon noise, which evidently has to be minimized.
The dimensions of the telescope are limited by the overall 
size of the satellite which in turn are constrained by the restrictions
of launch opportunities. 
This determines the length of the telescope;
as discussed in \secref{sec:orbit} the diameter is then
effectively limited by the constraint that the Earth exclusion angle
be at most 30\deg; this has led to a telescope diameter of 32~cm
and a telescope with an f-ratio of 0.9.

\begin{figure}[!htp]
\begin{center}
    \resizebox{8.8cm}{!}
    {\includegraphics{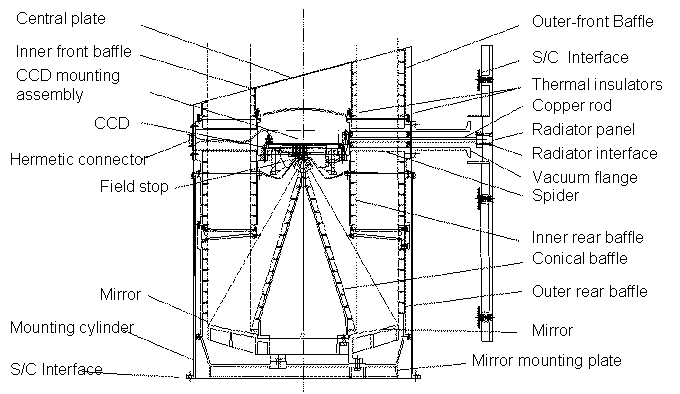}}
\end{center}

\caption{
Schematic design of the MONS Main Telescope.
The field stop is mounted on the conical baffle
which is bolted to the main mirror, to ensure mechanical and
thermal stability.
The CCD is cooled through a heat pipe connected to the radiator panel.
\label{fig:teldesign}
}
\end{figure}

Light will be detected by means of a CCD detector.
The total number of photons detected is therefore limited by
the well depth of the detector;
to reduce the photon noise the stellar image is defocused to an
outer diameter of 9~mm.
To reduce sensitivity to guiding errors the point-spread function
must be uniform to 1 -- 2 \% over small scales.
Effects of scattered light are minimized by passing light from the
star through a field stop, placed at the focal position, with a
diameter of less than 1~mm.
The field stop is mounted on a conical baffle directly bolted to the
primary mirror; mirror and baffle are made of aluminium, so that
temperature changes will not change the relative dimensions of
the optical system. 
The overall telescope design is shown in \figref{fig:teldesign}.

\begin{figure}
\begin{center}
    \resizebox{7.cm}{!}
    {\includegraphics{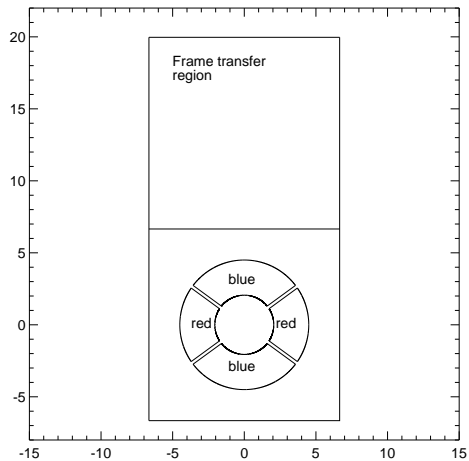}}
\end{center}

\caption{
Schematic illustration of the detector plane. 
The pupil image is divided into segments of blue and red light,
as indicated. 
The total size of the CCD detector is ${\rm 2k \times 1k}$ pixels,
of which the upper half is used for frame transfer.
\label{fig:detect}
}
\end{figure}

\begin{figure}
\begin{center}
    \resizebox{8cm}{!}
    {\includegraphics[bb=95 405 542 720]{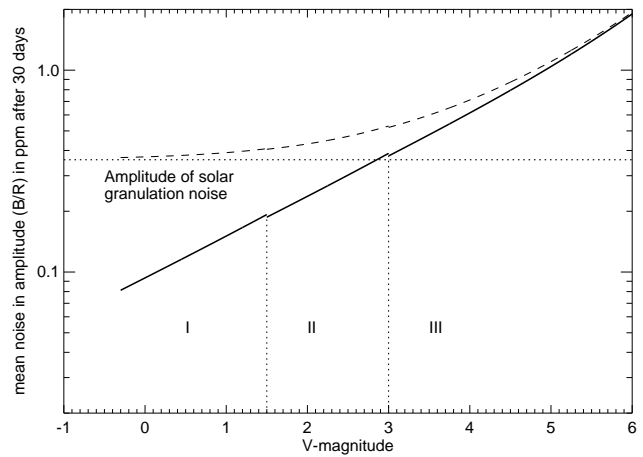}}
    \vskip 0.5cm
\end{center}

\caption{
The noise level in the colour-ratio observations, in parts per million,
for a 30-days observing sequence,
as a function of stellar magnitude.
The solid line shows the instrumental noise, the dotted line the
intrinsic noise at the solar level and the dashed line is the sum.
Note that the CCD is operated in three different modes
(I, II, and III), depending on the magnitude of the star.
\label{fig:mmtscatter}
}
\end{figure}

To reduce instrumental noise it is highly desirable to carry out
differential photometry.
This is achieved in the present design by measuring differentially
between light in two wavelength bands (red and blue) from the target star;
since the oscillation amplitude is substantially higher in the
blue waveband than in the red band ({\cf} Fig.~\ref{fig:VIRGO}),
the ratio between the two intensities contains an oscillation signal.
The separation into two bands is made by placing a filter,
divided into red and blue segments, at a suitable point in the optical path;
as a result, the two bands are measured simultaneously
on the same detector, ensuring the full advantage of the differential
measurement, although at the expense of some loss of light.
A schematic view of the detector plane is shown in Fig.~\ref{fig:detect};
note that the defocussed image fills a large fraction of the CCD.

The baseline CCD detector has an active area of ${\rm 1k \times 1k}$ pixels,
with frame-transfer readout.
It is cooled passively
to a temperature of $-90$~C,
by connecting the detector by means of a cold finger to a cooling plate.
The temperature will be kept stable to $0.1$~C through a small heating element
and measured to a precision of $0.01$~C for later decorrelation
during data analysis.
The expected noise level in the measurement of the colour ratio,
assuming a 30-day observing sequence, is illustrated in 
\figref{fig:mmtscatter}.
This satisfies the goal of reducing the measurement error to be
below the expected intrinsic stellar `noise' for the brighter targets. 

\subsection{MONS Field Monitor}
\label{sec:MFM}

The field stop of the MONS Main Telescope lets though light
from a region with a diameter of around 11 arcmin
on the sky.
Thus the defocused image on the detector will contain contributions
from all stars within such a region around the main target.
Large-amplitude variations of a faint source could be mistaken
for oscillations of the target.
Although it would in principle be possible to test against this
by means of contemporaneous ground-based observations, the
effort required to do so would in practice be unrealistic.
Therefore we have included a small focused telescope to monitor
the field around the target.
The requirements for this Field Monitor can be estimated by noting that
if a magnitude difference of at least $\Delta V = 6$ is assumed between
the target and the neighbouring stars, the required photometric
precision for the target translates into a noise level of at most
125 ppm in the amplitude spectrum after 30 days, in the
Field-Monitor observations of a $V = 10$ star.

\begin{figure}
\begin{center}
  \resizebox{6.cm}{!}
  {\includegraphics{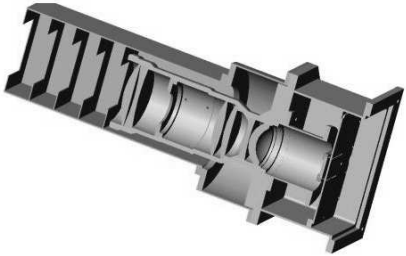}}
\end{center}

\caption{
Current design of the MONS Field Monitor.
Note the careful baffling.
\label{fig:mfm}
}
\end{figure}

The MONS Field Monitor consists of a lens assembly with an
aperture of 50~mm, with a ${\rm 1k \times 1k}$ 
CCD detector that gives a field of $5^\circ \times 5^\circ$.
Since the precision requirements are less stringent than for the
Main Telescope, so are the requirements on cooling of the detector;
it will be maintained at a temperature of around $-20$~C.
The current design is illustrated in Fig.~\ref{fig:mfm}.

\begin{figure}
\begin{center}
  \resizebox{8.cm}{!}
  {\includegraphics{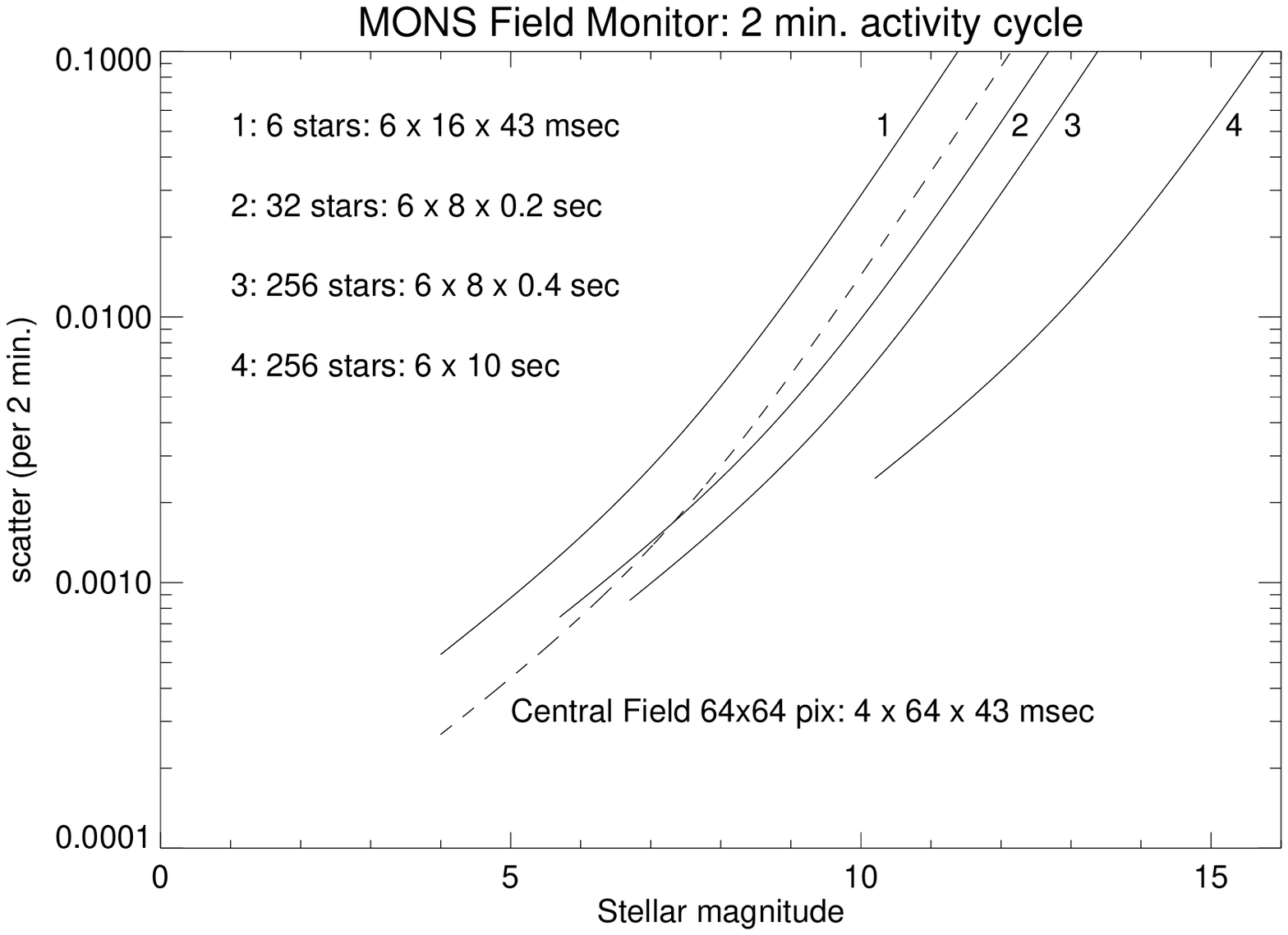}}
  \par
  \resizebox{8.cm}{!}
  {\includegraphics{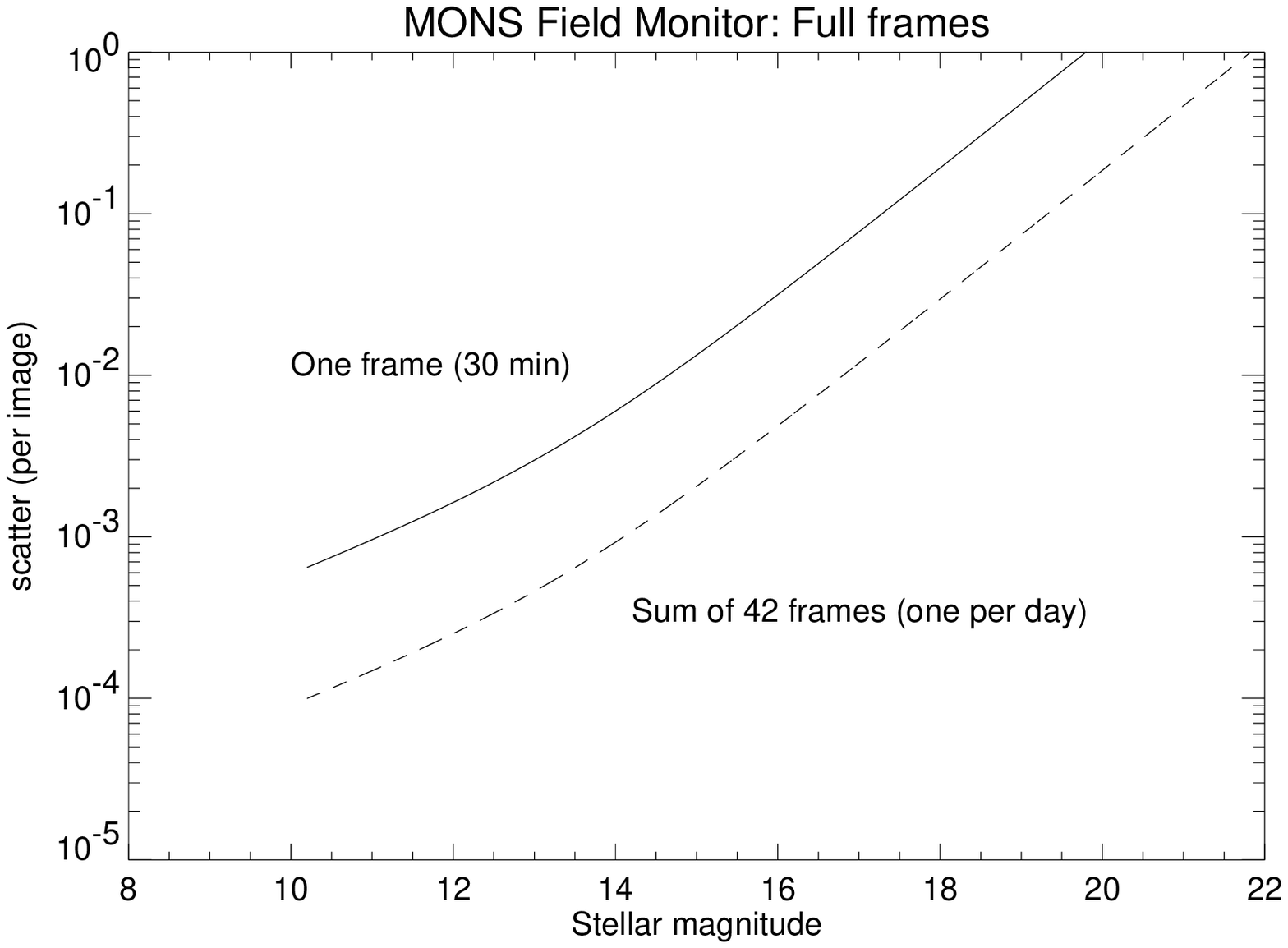}}
\end{center}

\caption{
Estimated photometric performance of the MONS Field Monitor,
as a function of $V$ magnitude.
The upper panel shows the rms scatter (in a 20-sec observing cycle)
for individual stars.
The dashed line shows the precision obtained in the central field,
used for correction of the main-target observations, which will
be analysed on the ground.
The solid lines show the precision for (from top to bottom)
integration times of 43 msec, 0.2 sec, 0.4 sec and 10 sec;
a possible distribution of the number of stars to be observed in
each cathegory, given constraints of on-board memory and telemetry,
is also indicated.
The lower panel shows the scatter per image in time-integrated
images over 30 min (solid line) and
summed over one day (dashed line).
\label{fig:mfm-perform}
}
\end{figure}

We note that the large focused field of the Field Monitor lends itself
to very interesting possibilities for parallel science.
In addition to the observations required for the primary science,
data on selected stars in the field will be recorded at a rapid
cadence, to investigate other types of pulsating stars.
Furthermore, the image of the complete field will be accumulated and
transmitted to the ground at a lower rate, at least several times
per orbit, to allow search for transient phenomena in a large number of objects.
The photometric capabilities of these observing modes are
illustrated in Fig.~\ref{fig:mfm-perform}.
Possible targets for parallel science are discussed in \secref{sec:parallel}.

\begin{figure}
\begin{center}
  \resizebox{4.3cm}{!}
  {\includegraphics{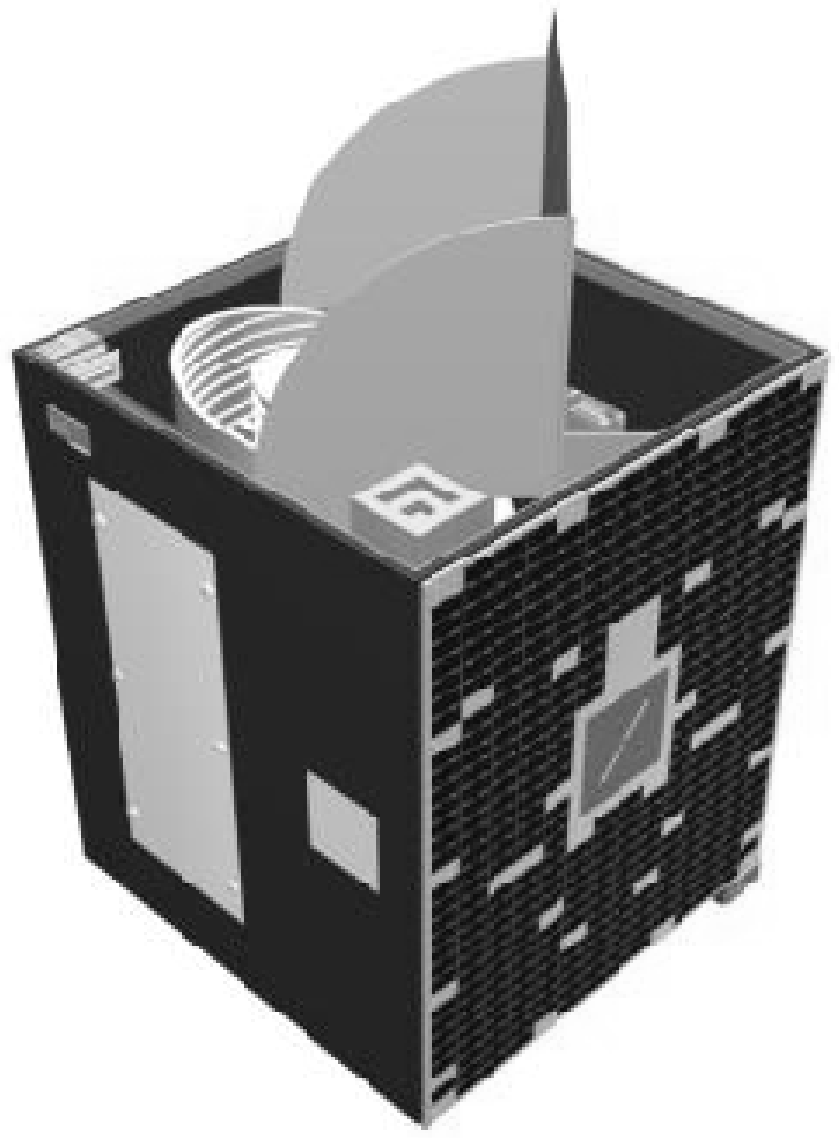}}
  \resizebox{4.3cm}{!}
  {\includegraphics{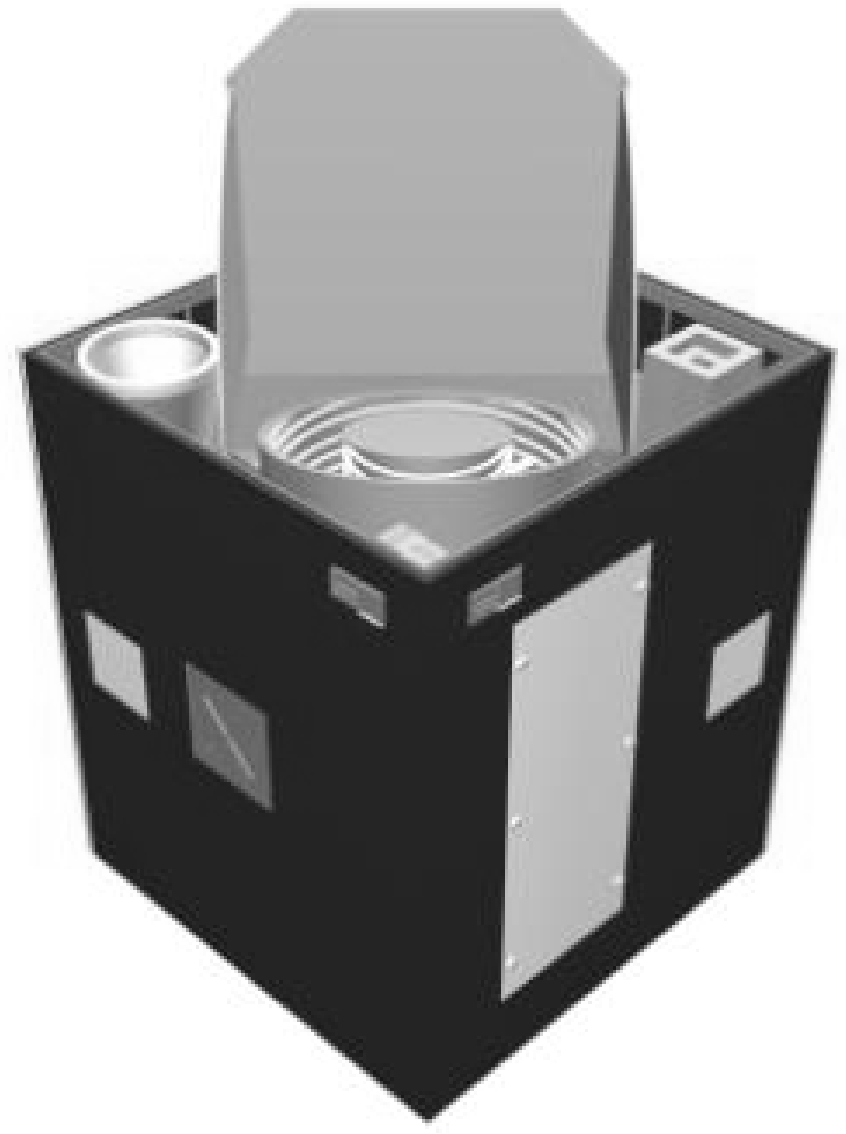}}
\end{center}

\caption{
Two different views of the R{\o}mer satellite conceptual design.
The left-hand image shows one sun-pointing side (covered with
solar panels and with one of the antennas) and one cold side;
note the (partly hidden) Main Telescope and the square baffle of
the Field Monitor.
The right-hand image shows the two cold sides, with radiator panels
to cool the CCD cameras; also visible is the round baffle of the
forward-pointing Star Tracker.
\label{fig:satview}
}
\end{figure}

\section{Platform}
\label{sec:platform}

The Systems Definition Phase of the R{\o}mer project led to the
conceptual design of the satellite illustrated in 
Fig.~\ref{fig:satview}. 
The overall dimensions are $60 \times 60 \times 71$~cm and the mass,
including a 25~\% margin, is around 100~kg,
well below the limit set by ASAP-5 constraints.
In operation,
the satellite will be oriented such that one diagonal points towards
the Sun; thus two sides are covered with solar panels, while the other two
sides, which are permanently cold, carry the cooling plates for
the CCD cameras.
The lid, which is closed during launch, protects the main telescope
against direct sunlight.
The solar cells allow an average power consumption of 55~W.

The satellite is three-axis stabilized,
controlled by four momentum wheels.
Momentum dumping will be made through magneto-torquing,
by means of magnetic coils coupling to the Earth's magnetic field,
during the parts of the orbit that are sufficiently close to the Earth.
Attitude control is maintained by means of two Star Trackers,
of which generally only one is operational at any given time.
Additional coarse attitude information is obtained from Sun sensors
and a magnetometer.
The expected attitude stability during observations is better than 10 arc~sec.

The platform is controlled by a single Command and Data Handling computer,
with a separate computer managing the payload and taking care
of the on-board data processing.
The various components are connected by a dual redundant CAN bus.
For communication the satellite carries two S-band antennas, 
with a transmission power of 2~W.
The baseline is to use a single ground station, located in Denmark,
with a 1.8-m antenna;
this allows a average data rate of 24 Mbyte/day, sufficient
for both the primary and parallel science.

\section{Orbit and operations}
\label{sec:orbit}

The targets for a modest asteroseismic mission like MONS are generally
bright stars, for which the most precise measurements can be made.
This furthermore has advantages in terms of the determination of
other properties of the stars: bright main-sequence stars are relatively
nearby and hence have accurate determinations of their distances;
and because they are bright, their spectroscopic properties can be
determined in great detail.

The disadvantage of the restriction to bright stars is that the
{\it selection} of stars is somewhat limited.
This is particularly constraining since, in addition, it is important
to observe as wide a range of stellar properties as possible,
as discussed in more detail in Section~\ref{sec:targets}.
Thus it is highly desirable to organize the mission, including most
importantly the choice of orbit, such that all parts of the sky can
be observed at some time during the mission.
Needless to say, this has to be balanced against the constraints 
implicit in a small-satellite mission, in terms of cost and operational
simplicity.

{}From a scientific point of view, the requirements are
that the relevant targets must be observable at least for one 30-day period
during the baseline two-year mission,
the observations being, as far as possible, without substantial
interruptions.
Also, the effects of scattered light from the Sun and Earth (and, to
lesser extent, the Moon) on the observations with the Main Telescope
must be minimized.
In addition, a number of technical constraints must be satisfied.
A part of the orbit must go through a region of the Earth's
magnetic field which allows momentum dumping with magnetorquers.
This requires a low perigee and a sufficiently high orbit inclination.
Also, part of the orbit must be close to the Earth, preferably within view of
Denmark, to simplify data transmission and control of the satellite.
The orbit must provide sufficient solar radiation on the solar panels
during the observing modes.
The use of a Star Tracker must be possible, without interference
from the Sun or Earth, for all the relevant targets.
And finally,
the constraints on cost clearly mean that an orbit must be chosen
where the satellite can be launched as a passenger.

In practice, the constraints on the Earth mean that
light from the Earth should never directly reach the main mirror
of the MONS telescope during observations.
Similarly, for the Star Trackers the Earth should be outside
an exclusion angle which depends on the length of the baffle
and other design aspects.

\begin{figure}
\begin{center}
    \resizebox{8cm}{!}
    {\includegraphics{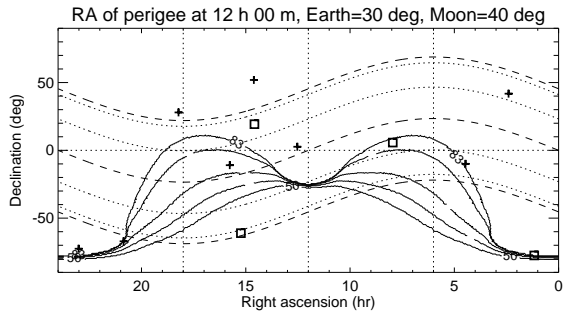}}
\end{center}

\caption{
\label{fig:excl}
The region of the sky that is excluded by Earth light on the main mirror
(limb of the Earth $30^\circ$ away from telescope axis). Also shown are
regions where the Moon can get within 40 degrees of the telescope axis.
The contours shows the duty cycle for a given position in the sky taking
into account the position of the Earth, the 
right ascension of the orbit perigee and
the radiation belts around the Earth. 
The maximum duty cycle is around 85~\% 
(limited by passage through the radiation belts).
The squares show the location of priority 1 targets and the plusses
the location of priority 2 targets ({\cf} \tabref{tab:targets}).
}
\end{figure}

Based on these considerations, the baseline orbit has been chosen
to be a so-called Molniya orbit, whose 
main characteristics are summarized in Table~\ref{tab:molniya};
such orbits are
used for Russian communication satellites, and hence launch
opportunities arise at regular intervals.
It is characteristic of the orbit that apogee is at high northern
latitude.
Thus light from the Earth is potentially troublesome for observations
of objects on the southern sky.
Following a trade-off analysis,
dimensions of the Main Telescope have been chosen such that
the Earth exclusion angle is $30^\circ$ ({\cf} \secref{sec:MMT}).
Given this, the possible sky coverage during the mission can
be determined, depending on the time of launch.
An example is illustrated in Fig.~\ref{fig:excl};
here also is shown the location in the sky of the highest-priority
targets for the main telescope ({\cf} Section \ref{sec:targets}).
Due to the precession of the orbit, the exclusion region shifts
through about 7 hours in right ascension during the two-year mission. 
Thus, with a suitable choice of the time of launch 
it will be possible to observe all key objects during the baseline mission.

Given the exclusion angle of $50^\circ$ of the Star Trackers,
observation of southern objects with just one forward-pointing
Star Tracker would be essentially impossible.
Thus the baseline satellite configuration has two Star Trackers,
pointing in the same and the opposite direction as the Main Telescope.
Only one of the Star Trackers need be operated at any given time for the
primary science programme, although the parallel science may benefit from
having both operating.

\begin{table}
\caption{Some properties of the Molniya orbit.
\label{tab:molniya}}
\begin{center}
\begin{tabular}{ll}
\hline
\noalign{\smallskip}
Launch		       & SOYUS/FREGAT \\
Period                 & 43080 sec (11.97 hrs)\\
Semi-major axis        & 26560 km\\
Radius of perigee      & 6900 km ($h \approx$ 500 km)\\
Radius of apogee       & 46200 km \\ 
Eccentricity           & 0.7411 \\
Inclination            & 63.43\deg \\
Argument of perigee     & $\sim$270\deg \\
Change in ascending node  & $-$0.150\deg/day\\
\noalign{\smallskip}
\hline
\end{tabular}
\end{center}
\end{table}

Passage through the Earth's radiation belts 
is an unavoidable consequence of the requirements of a high
apogee and a low perigee.
For the Molniya orbit this effect is to some extent reduced by the
high inclination, but not eliminated.
This results in requirements
on the radiation shielding which can, however, readily be met within
the mass budget of the satellite.
Another consequence is that the observations are interrupted 
during passage through the radiation belts. 
We estimate that the resulting duty cycle is around 85~\%;
except for very special cases 
this has minimal consequences for the quality of the data, compared
with a duty cycle of 100~\%.

In planning the observations, the minimum time block is 1 orbit;
this provides
around 10.5 hr useful observations out of each 12 hours.
The baseline operations schedule of the mission is very simple:
for most targets the demands of frequency resolution and
signal-to-noise ratio require extended and nearly continuous
observations of each target.
Thus the baseline is to observe each target for at least 30 days,
or 60 orbits.
Selected particularly important targets may be observed for 100 orbits.


For particular targets considerably shorter observing periods
(although always at least one orbit) may be used.
These include eclipsing binaries and planet transits;
here the precise time of the relevant event can be predicted 
well in advance, and the duration of the event is typically
less than one orbit.
Such observations will be scheduled, as far as possible, between
the main long-duration pointings.
However, in exceptional cases it may be possible to interrupt
the long pointings for a single orbit, without substantial effect
on the data quality.

\section{Preliminary target list}
\label{sec:targets}

When assigning priorities to solar-like targets for the MONS Main
Telescope, consideration must be given to the expected signal-to-noise,
as well as to
the desire to cover an interesting range in stellar parameters.
The relevant stellar properties include mass, evolutionary stage,
content of heavy elements (conventionally known as metals),
and rotation rate.
In some cases identification of stars with the desired property,
while still observable with \MONS, presents some difficulties;
in particular, metal-poor stars are fairly rare and therefore tend
to be rather distant and hence difficult to observe.
Also, it may happen that a star is potentially very interesting, but of 
uncertain observability.
In such cases further studies, either from the ground or with brief
observations with the \MONS{} Main Telescope, may be required for the
final selection.

Target selection has been based on an extensive list of stars
of sufficient brightness to be observable with \MONS{} and of the
appropriate spectral type.
These were then analysed in terms of their properties, as well as
for the expected amplitude and observational noise,
and discussed in the MONS Principal-Investigator team.
The result was the preliminary list
of targets presented in Table~\ref{tab:targets}, with stars grouped according
to properties and divided according to the following priorities: 

\begin{list}{}
{\setlength{\leftmargin}{20pt}
\setlength{\labelwidth}{20pt}}

\itemsep=0pt

\item[1.]  Very high priority.  Should definitely be observed.

\item[2.]  High priority.  Should be observed.

\item[2b.]  High priority but may not be feasible.  Should be observed if
feasible.

\item[3.]  Excellent target.  A small subset of this group will be observed

\item[3b.]  Excellent target but may not be feasible.  Should be added to
Priority 3 list if feasible.

\end{list}
The 2b and 3b groups contain stars which are very interesting, but which
may not show solar-like oscillations because of S/N limitations, or because
they are too hot.  These stars could be checked out for one or two orbits
to determine feasibility.

\begin{table*}
\caption[]{\label{tab:targets}
Possible solar-like targets for the MONS Telescope}
\leftskip=60pt
\smallskip
\begin{tabular}{lcccccccr}

\hline
\hline
\noalign{\smallskip}
   & Solar-         & Low-             & Higher-      & Metal-poor   & Metal-     & Fast         & Hotter      & Total \\
   &       mass     &     mass         &        mass  &              &       rich &      rotation&             &       \\
\noalign{\smallskip}
\hline
\hline
\noalign{\smallskip}
1. & $\alpha$~Cen~A & $\alpha$~Cen~B   & $\eta$~Boo  &               &            &              &              & 4\rlap{$^*$} \\
   & $\beta$~Hyi    &                  & $\alpha$~CMi \\
\noalign{\smallskip}
\hline
\hline
\noalign{\smallskip}
2. & $\mu$~Her      &                  & $\upsilon$~And& $\nu$~Ind   &$\delta$~Pav& $\theta$~Boo &              & 7 \\
   &                &                  & $\beta$~Vir  &              &            &              &              &   \\
\noalign{\smallskip}
\hline
\noalign{\smallskip}
2b.&                & $\varepsilon$~Eri&              & HD~140283    &            &              &              & 2 \\
\noalign{\smallskip}
\hline
\hline
\noalign{\smallskip}
3. & $\delta$~Eri   &                  & $\zeta$~Her  & $\gamma$~Pav &            & $\psi$~Cap   & $\alpha$~Tri &11 \\
   & $\eta$~Cas     &                  & $\alpha$~For &              &            &              &              &   \\
\noalign{\smallskip}
\hline
\noalign{\smallskip}
3b.&                & $\tau$~Cet       &              &              &            & $\kappa$~Cet & $\gamma$~Dor & 9 \\
   &                & 70~Oph           &              &              &            & $\chi^1$~Ori &             &   \\
   &                & 36~Oph           &              &              &            &              &              &   \\
\noalign{\smallskip}
\hline
\hline
\noalign{\smallskip}
%
\noalign{\smallskip}
\noalign{$^*$Note that $\alpha$~Cen~A and~B will be observed simultaneously.}

\end{tabular}

\end{table*}

The present target list is based on current information about the
stars, as well as on our present understanding of the excitation
of the oscillations.
The stars of priority~1 will certainly be observed; apart from
their intrinsic scientific interest observations from the ground
and from the WIRE satellite have demonstrated that they show
solar-like oscillations at an amplitude sufficiently high to
enable very detailed studies.
The final selection of the remaining targets will depend on
further studies, including ground-based observations to evaluate
the fields and the properties of the stars, as well as on modelling
to investigate the extent to which the observations may be expected
to provide the required information about the stellar properties.
Although a definite programme for the start of the mission will
evidently be established well before launch, adjustments of the
later programme will be possible, in the light of the information
obtained from the first observations.
Evidently, information about the general properties of solar-like
oscillations obtained with the MOST mission
({\cf} Matthews, these proceedings)
expected to be launched well before R{\o}mer, will also be taken
into account.

The stars listed in \tabref{tab:targets} may loosely be characterized
as solar-like oscillators.
In addition to these, a modest amount
of observing time will be assigned to other types of pulsating stars
as well as to studies of eclipsing binaries and planet transits.

\section{Parallel science}
\label{sec:parallel}

As discussed by Gilmore (these proceedings)
space-based photometry offers unique possibilities, for a broad
range of astrophysical objects and phenomena, in terms
of precision and continuity.
Although the R{\o}mer instrumentation is more modest than
what will be provided by the Eddington mission discussed by Gilmore,
the nearly continuous observations of fairly extended fields,
by the Field Monitor and the Star Trackers, 
allow potentially very valuable studies within a number of areas.

The potential of observations from 
space of even modest instrumentation was demonstrated by
the 5.4-cm star tracker on the otherwise failed WIRE satellite,
which has provided valuable data on low-amplitude stellar
pulsation, despite the rather unsuitable orbit of WIRE
({\eg} Buzasi {\etal} 2000; Cuypers, these proceedings). 
The R{\o}mer Star Trackers will have a precision better than
3~mmag in one minute,
for all stars brighter than magnitude $V=6$.
As illustrated in Fig.~\ref{fig:mfm-perform}
an even better performance is expected for the Field Monitor.
As a result, science with the Star Trackers 
and the Field Monitor is seen as a high-priority
goal by a broad international community,
and must be considered as an important part of the mission,
in parallel with the science to be carried out with the MONS Main Telescope.


Potential targets for the parallel science 
include `classical' pulsating stars;
for these we expect a substantial increase in the
number of modes detected, compared with ground-based results,
possibly leading to the identification of a fairly substantial
fraction of the modes within the part of the oscillation spectrum,
typically a frequency range, where unstable modes are found.
As a result we may hope, from the frequency patterns, to be able
to determine the nature of the modes observed; this is a prerequisite
for the full use of the frequencies for asteroseismology.
Our inability so far to identify the modes has been a major
impediment to the investigations of stellar interiors from
observed frequencies of `large-amplitude' pulsating stars.

In several cases, particularly solar-like oscillations in subgiants,
we expect the Star Trackers to be sufficiently sensitive to determine
at least the overall properties of the oscillations of the stars.
In such cases, Star-Tracker observations can serve as very useful
precursors for later observations of a star with the Main Telescope:
from the determination of the amplitude and gross frequency characteristics
we shall be in a better position to optimize the observing programme
to be carried out.

Summed images, in particular from the Field Monitor, have a very
interesting potential for studying variations on longer time scales
of faint objects.
Examples include detection of extra-solar planets through transits, 
detection of supernova explosions in distant galaxies and
the possible detection of smaller objects in our solar system.

We also note that the data from the Star Trackers have very considerable
potential for use in outreach activities, such as projects for
schools or amateur astronomers.

\section{The MONS Scientific community}
\label{sec:MSC}

The MONS project involves a substantial international community,
comprising around 150 scientists, taking
part both in the preparation of the mission and in the planning
of the utilization of the data.
This is organized in the {\em MONS Science Consortium}.
Detailed information about the activities of the MSC can
be found on {\tt http://astro.ifa.au.dk/MSC}).
Within the MSC several working groups have been established
to deal with specific aspects of the project, such as different
types of targets or the ground-based support observations:

\begin{itemize}

\item {\bf Solar-like oscillations}:
This group deals with stars showing solar-like oscillations, 
{\ie}, modes excited stochastically by convection. 
These are typically stars on or near the main sequence,
which will be targets for the MONS Main Telescope. 

\item {\bf B stars}:
This group deals with pulsating B stars, including $\beta$ Cephei stars,
slowly pulsating B stars and Be stars. 
These will in most cases be observed with the Star Trackers
and Field Monitor,
but a few may be selected for observation with the Main Telescope. 

\item {\bf A and F stars}:
This group will deal with pulsating A and F stars, 
including $\delta$ Scuti stars and rapidly oscillating Ap stars. 
They will in most cases be observed with the Star Trackers
and Field Monitor,
but a few may be selected for observation with the Main Telescope. 

\item {\bf Planets and eclipsing variables}:
This group deals with study of eclipsing binaries and transits 
of giant planets. 
These will in most cases be observed with the Star Trackers
and Field Monitor, but particularly interesting, predicted,
transits could be selected for observation with the Main Telescope. 

\item {\bf Ground-based support observations}:
This group takes care of the ground-based observations required
for the MONS project, including redetermination of stellar
parameters, investigations of target fields and follow-up
observations of pulsations ({\eg} to help mode identification).

\end{itemize}

The primary MONS data analysis will be carried out at the Science Data Centre
at the University of Aarhus.
Distribution of data and other relevant information, such as
results of ground-based support observations and theoretical results,
will be organized through the MONS Information System
(see {\tt http://astro.ifa.au.dk/MIS}).

\section{Schedule}
\label{sec:sched}

The Detailed Design Phase of the R{\o}mer project was formally started
at a kick-off meeting on 2 October 2001.
This phase is funded by a grant of 12 MDkr (around 1.5 MEuro) from
the original phase of the Danish Small Satellite Programme.
Complete funding is being sought at the time of writing through 
extension of the programme beyond 2001.
Assuming that funding is made available at the start of 2002,
the Implementation Phase will commence in the middle of 2002,
while the Detailed Design Phase will end by a Detailed Design Review
at the end of 2002.
With this schedule launch is foreseen in the middle of 2005;
however, a definite schedule will have to await 
the establishment of a more precise funding profile.


\begin{acknowledgements}

I am very grateful to T.~R.~Bedding and H.~Kjeldsen for extensive 
material for this overview, as well as for a very enjoyable and
fruitful collaboration on the MONS project.
I thank B.~L.~Christensen-Dalsgaard for help with the intricacies
of image conversion.
This work was supported in part by the Danish National Research 
Foundation through the establishment of the Theoretical
Astrophysics Center.

\end{acknowledgements}


\begin{thebibliography}{}

\bibitem[]{}
Bedding T. R., Butler R. P., Kjeldsen H., Baldry I. K., 
O'Toole S. J., Tinney C. G., Marcey G. W., Kienzle F., Carrier F. 2001,
ApJ {\rm 549}, L105 
\bibitem[]{}
Bouchy F., Carrier F. 2001,
A\&A {\rm 374}, L5 
\bibitem[]{}
Brown T. M., Gilliland R. L. 1994,
ARAA {\rm 32}, 37 
\bibitem[]{}
Buzasi D., Catanzarite J., Laher R., Conrow T., Shupe D.,
Gautier III T. N., Kreidl T., Everett D. 2000,
ApJ {\rm 532}, L133 
\bibitem[]{}
Christensen-Dalsgaard J. 1998, 
in `Proceedings of Workshop on Science with a Small Space telescope',
eds Kjeldsen H., Bedding T.R., Aarhus University, 17 
\bibitem[]{}
Christensen-Dalsgaard J., Frandsen S. 1983,
{\rm Solar Phys.} {\rm 82}, 469 
\bibitem[]{}
Christensen-Dalsgaard J., D\"appen W., Dziembowski W. A.,
Guzik J. A. 2000,
in `Variable Stars as Essential Astrophysical Tools',
ed. \.{I}bano\u{g}lu C., Kluwer Academic Publishers, p. 59 
\bibitem[]{}
Fr\"ohlich C., Crommelynck D. A., Wehrli C., Anklin M.,
Dewitte S., Fichot A., Finsterle W., Jim\'enez A.,
Chevalier A., Roth H. 1997,
{\rm Solar Phys.} {\rm 175}, 267 
\bibitem[]{}
Gautschy A., Saio H. 1996,
ARAA {\rm 34}, 551 
\bibitem[]{}
Gough D. O. 1998,
in `Proceedings of Workshop on Science with a Small Space telescope',
eds Kjeldsen H., Bedding T.R., Aarhus University, p. 33 
\bibitem[]{}
Houdek G., Balmforth N. J., Christensen-Dalsgaard J., Gough D. O. 1999,
A\&A {\rm 351}, 582 
\bibitem[]{}
Kjeldsen H., Bedding T. R. 1995,
A\&A {\rm 293}, 87 
\bibitem[]{}
Kjeldsen H., Bedding T. R., Viskum M., Frandsen S. 1995,
AJ {\rm 109}, 1313 
\bibitem[]{}
Marti\'c M., Schmitt J., Lebrun J.-C., Barban C., Connes P.,
Bouchy F., Michel E., Baglin A., Appourchaux T., Bertaux J.-L. 1999,
A\&A {\rm 351}, 993 
\bibitem[]{}
Monteiro M. J. P. F. G, Christensen-Dalsgaard J., Thompson M. J. 2000,
MNRAS {\rm 316}, 165 
\bibitem[]{}
Olsen N., Holme R., Hulot G., {\etal} 2000,
{\rm Geophys. Res. Lett.} {\rm 27}, 3607 
\bibitem[]{}
P\'erez Hern\'andez F., Christensen-Dalsgaard J. 1998,
MNRAS {\rm 295}, 344 
\bibitem[]{}
Soufi F., Goupil M. J., Dziembowski, W. A. 1998,
A\&A {\rm 334}, 911 

\end{thebibliography}
\end{document}